\documentclass[12pt, twoside]{article}
\usepackage{a4wide,amssymb,cite}
\usepackage{epsfig}

\newcommand{\be}{\begin{equation}}
\newcommand{\ee}{\end{equation}}
\newcommand{\bea}{\begin{eqnarray}}
\newcommand{\eea}{\end{eqnarray}}

\begin{document}

\begin{titlepage}

\rightline{CERN-PH-TH/2010-102}

\begin{centering}
\vspace{1cm}
{\large {\bf  Chaotic inflation in Jordan frame supergravity}}\\

\vspace{1.5cm}

 {\bf Hyun Min Lee}
\\
\vspace{.2in}

{\it CERN, Theory division, CH-1211 Geneva 23, Switzerland} \\

\vspace{.1in}
(e-mail address: hyun.min.lee@cern.ch)

\end{centering}
\vspace{2cm}

\begin{abstract}
\noindent
We consider the inflationary scenario with non-minimal coupling in 4D Jordan frame supergravity.
We find that there occurs a tachyonic instability along the direction of the accompanying non-inflaton field in generic Jordan frame supergravity models. We propose a higher order correction to the Jordan frame function for solving the tachyonic mass problem and show that the necessary correction can be naturally generated by the heavy thresholds without spoiling the slow-roll conditions. We discuss the implication of the result on the Higgs inflation in NMSSM.

\end{abstract}

\vspace{1cm}
%



\end{titlepage}

\section{Introduction}

Cosmic inflation \cite{inflation} has been a paradigm beyond the Standard Big Bang Cosmology in which flatness, isotropy, homogeneity, horizon and relic problems are explained and solved. The inflaton generates scale-invariant and Gaussian spectrum of density fluctuations. Furthermore, quantum fluctuations during inflation provides a seed for the large-scale structure formation as we see now.

An economical proposal to utilize the Higgs doublet in the Standard Model(SM) as the inflaton has recently drawn some attention \cite{higgsinf}. This is the so called Higgs inflation. In this scenario, the chaotic inflation can be realized due to a large non-minimal coupling of the Higgs doublet to gravity \cite{nonminimal}, instead of having a tiny Higgs quartic coupling, which is contradictory with the Higgs mass bound.
However, some time after the proposal, it has been shown by power counting formalism that the Hubble scale during inflation is proximate to the unitarity bound on the new physics scale associated with the breakdown of the semi-classical approximation in the effective theory \cite{boundH}.
This result is independent of the frames and the backgrounds that the power counting is applied to \cite{frameindep}. 
Nonetheless, a singlet field with non-minimal coupling could be a viable inflaton candidate for a small singlet quartic self-coupling for which the Hubble scale can be smaller than the unitarity bound.

Weak-scale supersymmetry \cite{susy} is a solution to the hierarchy problem in the SM
and has been one of main topics in the search for new physics at the Large Hadron Collider(LHC).
In the Minimal Supersymmetric Standard Model(MSSM), there are two Higgs doublets and the Higgs quartic coupling is given in terms of the electroweak gauge couplings. 
Then, one may ask whether SUSY can help address the naturalness issue of the Higgs inflation in the context of the MSSM. 
However, apart from the unitarity problem in the Higgs inflation, it has been shown that in the MSSM, 
the Higgs inflation cannot be realized due to the instability along the $\beta$ field which is the ratio of two Higgs VEVs \cite{higgsnmssm}.

On the other hand, in the Next-to-Minimal Supersymmetric Standard Model(NMSSM) \cite{nmssm}, 
an additional Higgs self-coupling can be introduced by the superpotential term coupling the Higgs doublets to a singlet chiral superfield and it can provide the vacuum energy needed for inflation \cite{higgsnmssm}.
Since this new Higgs self-coupling can be made small without violating the LEP bound on the Higgs mass,
there is a possibility for the Higgs inflation to work within the semi-classical approximation.
However, even in this case, the singlet field, which would be a non-inflaton field, gets a tachyonic mass during inflation and would spoil the slow-roll inflation of the Higgs fields \cite{jsugra}.

In this paper, we revisit the tachyonic mass problem in the NMSSM in 4D Jordan frame supergravity \cite{jsugra}.
For this purpose, we consider a simple toy model with two singlet chiral superfields\footnote{The single field inflation with non-minimal coupling cannot be realized in supergravity.} to capture the main difficulty of the inflation with non-minimal coupling in supergravity and provide a solution to the problem.  Thus, we introduce two singlet fields: one singlet field becomes the inflaton and the other singlet field provides a nonzero F-term potential through the coupling to the inflaton field.
Then, we find that there appears a tachyonic instability along the non-inflaton singlet for the minimal form of the frame function due to the negative supergravity mass correction.

As a solution to the tachyonic mass problem, we add a higher order correction for the non-inflaton field in the frame function. In this case, for an appropriate value of the coefficient of the new term, 
we show that it is possible to make the non-inflaton singlet field get a positive squared mass and stable during inflation while the inflaton dynamics is unchanged. We give an example where heavy fields coupled only to the non-inflaton field
generates such a higher order correction with necessary coefficient in the one-loop effective frame function.
Our result can be applied directly to the Higgs inflation with zero D-term in the NMSSM.
For a successful Higgs inflation in the NMSSM, the Higgs parameters in the NMSSM are constrained for the necessary inflationary parameters. In particular, the non-minimal Higgs coupling gives rise to the effective $\mu$ term by Giudice-Masiero mechanism \cite{gmmech}. As a result, for a large non-minimal coupling, gravitino mass is much smaller than the effective $\mu$ term which is of order the soft mass parameters. Thus, gravitino can be LSP and become a dark matter candidate.

The paper is organized as follows.
We first explain a general framework for 4D Jordan frame supergravity where non-minimal couplings for scalar fields are suitably introduced.
Then we take a minimal inflationary model with two singlet chiral superfields
and point out the tachyonic mass problem.
Consequently, we propose a solution to the tachyonic instability problem and find a necessary condition for satisfying the slow-roll inflation and the unitarity bound on the heavy field mass.
Next we discuss the implication of the Higgs inflation on the NMSSM phenomenology in the later section.
Finally, a conclusion is drawn.
There are two appendices dealing with the K\"ahler metric in Jordan frame supergravity and 
containing an example where the one-loop correction to the frame function is calculated in the presence of heavy fields.

\section{Jordan frame supergravity}

We start with the general Einstein-frame action in 4D ${\cal N}=1$ supergravity \cite{esugra},
\be
S_E=\int d^4 x \sqrt{-g_E}\Big(\frac{1}{2}R -K_{i{\bar j}}D_\mu\phi^i D^\mu {\bar\phi}^{\bar j}-V_E\Big)
\ee
where the covariant derivatives for scalar fields $\phi^i$ are given by $D_\mu\phi^i=\partial_\mu\phi^i-A^a_\mu \eta^i_a$.
Here the Einstein-frame scalar potential is given in terms of the K\"ahler potential $K$, the superpotential $W$ and the gauge kinetic function $f_{ab}$ by
\be
V_E = V_F + V_D \label{spot}
\ee
where
\bea
V_F &=& e^K \Big((K^{-1})^{i{\bar j}}(D_iW)(D_{\bar j} W^\dagger)-3|W|^2\Big), \\
V_D &=& \frac{1}{2} {\rm Re}f^{-1}_{ab} \Big(-i\eta^i_a\partial_i K +3ir_a\Big)\Big(-i\eta^i_b\partial_i K +3ir_b\Big)
\eea
with $G\equiv K+\ln |W|^2$ and the gauge transformations of the K\"ahler potential and the superpotential being $\delta_a K=3(r_a+{\bar r}_a)$ and $\delta_a W=-3r_a W$, respectively. We note that $r_a$ is nonzero only for the gauged $U(1)_R$ symmetry for which the superpotential transforms with $r_R=-\frac{2}{3}ig_R$. 

Performing a Weyl transformation of the metric with $g^E_{\mu\nu}=(-\Omega/3)g^J_{\mu\nu}$, we obtain the general Jordan-frame supergravity action from the above Einstein-frame action as follows,
\be
S_J=\int d^4x \sqrt{-g_J}\Big(-\frac{1}{6}\Omega R-\frac{1}{4\Omega}(\partial_\mu\Omega)(\partial^\mu\Omega)
+\frac{1}{3}\Omega K_{i{\bar j}}D_\mu\phi^i D^\mu {\bar\phi}^{\bar j}-V_J  \Big)
\ee
where the Jordan-frame scalar potential is related to the Einstein-frame one as
\be
V_J = \frac{\Omega^2}{9} V_E. \label{jspot}
\ee
The complete Jordan-frame supergravity action including fermions and gauge bosons can be found in Ref.~\cite{jsugra}.
Now specifying the frame function $\Omega$ to the K\"ahler potential as
\be
\Omega = - 3 M^2_P e^{-K/(3M^2_P)}, \label{framefunc}
\ee
we simplify the Jordan-frame action \cite{jsugra} as
\be
S_J=\int d^4x \sqrt{-g_J}\Big(-\frac{1}{6}\Omega R-\Omega_{i{\bar j}}D_\mu\phi^i D^\mu {\bar\phi}^{\bar j}+\Omega b^2_\mu-V_J  \Big)
\ee
where the auxiliary vector field $b_\mu$ gets the following form,
\be
b_\mu =\frac{1}{2}iA^a_\mu(r_a -{\bar r}_a) -\frac{i}{2\Omega}\Big(D_\mu\phi^i\partial_i\Omega - D_\mu {\bar\phi}^{\bar i}\partial_{\bar i}\Omega\Big). \label{auxvec}
\ee
Therefore, the kinetic terms for scalar fields depend on the analogue of the K\"ahler metric with $\Omega$ playing a role of $K$. However, the geometry of the non-linear sigma model of scalar is not of the K\"ahler type because of the additional term proportional to $b^2_\mu$.

In order to get the canonical scalar kinetic terms in the Jordan frame, we need $\Omega_{i{\bar j}}=\delta_{i{\bar j}}$ and $b_\mu=0$.
The most general frame function for giving $\Omega_{i{\bar j}}=\delta_{i{\bar j}}$ is the following \cite{jsugra},
\be
\Omega= - 3M^2_P + \delta_{i{\bar j}} \phi^i {\bar\phi}^{\bar j} -\frac{3}{2}(F(\phi)+{\rm h.c.}).
\ee
Then, from the relation (\ref{framefunc}), the corresponding K\"ahler potential takes the following form,
\be
K= - 3M^2_P \ln \Big(1-\frac{1}{3M^2_P}\delta_{i{\bar j}} \phi^i {\bar\phi}^{\bar j} +\frac{1}{2M^2_P}(F(\phi)+{\rm h.c.})\Big).
\ee  
Even with this choice of the frame function, we note that the auxiliary vector field $b_\mu$ is nonzero due to the angular modes of complex scalar fields.
During the cosmological evolution, however, when only the moduli $|\phi^i|$ dominate the dynamics, the scalar kinetic terms can be of canonical form. When $F=0$, the non-minimal coupling of the scalar fields are fixed as ${\cal L}=-\sqrt{-g}\,\sum_i\xi_i|\phi_i|^2R$ with $\xi_i=\frac{1}{6}$ so the scalar fields are conformally coupled to gravity. 
However, by choosing an appropriate holomorphic function $F$, we can break the conformal symmetry explicitly and include the nontrivial non-minimal coupling to gravity. 
Thus, it is possible to get a supergravity realization of the inflation model with non-minimal coupling. 
Henceforth we set the Planck scale to $M^2_P=1$ but we will recover $M_P$ whenever needed.

\section{The tachyonic mass problem in Jordan-frame supergravity inflation}

We consider an inflation model with two singlets $S$ and $X$ in the Jordan frame supergravity.
For the canonical scalar kinetic terms with two singlets, the general frame function in the Jordan frame is
\be
\Omega=-3+S^\dagger S + X^\dagger X -\frac{3}{2}(F(S,X)+{\rm h.c.}). \label{framefunc1}
\ee
Now we choose a non-minimal coupling to be $F=\chi S^2$ with $\xi$ being a dimensionless constant.
Then, the K\"ahler potential becomes
\be
K= - 3\ln \Big(1-\frac{1}{3}S^\dagger S -\frac{1}{3}X^\dagger X +\frac{1}{2}(\chi S^2+{\rm h.c.})\Big).
\ee
Moreover, by imposing a $U(1)_G$ global symmetry\footnote{The R-symmetry only does not restrict the superpotential to the form considered in this paper, rather also allowing for a tadpole term $W=f^2 X$, which would affect the slow-roll inflation
unless $|f|\ll \frac{|\lambda|}{\sqrt{2}}|S|$. We allow only for the dimensionless coupling by imposing the non-R $U(1)_G$ symmetry, which is the analogue of PQ symmetry in the NMSSM.} with charges, $G[X]=-2$ and $G[S]=+1$, we find the following unique superpotential at the renormalizable level,
\bea
W= \frac{1}{2}\lambda X S^2. 
\eea
where $\lambda$ is a dimensionless coupling.
The first attempt for the single field inflation with $W=\frac{1}{3}\lambda S^3$ was unsuccessful due to the negative vacuum energy coming from the supergravity correction \cite{higgsnmssm}. This is in a similar spirit to the general problem in supergravity chaotic inflation models with a single field \cite{chaoticsugra}.
We note that the $U(1)_G$ global symmetry is broken explicitly in the K\"ahler potential only by the non-minimal coupling. 

With no gauged $U(1)_R$ symmetry, the scalar potential for the singlets comes only from the F-terms and it is given in the Jordan frame by using eq.~(\ref{jspot}) with eq.~(\ref{spot}),
\bea
V_J&=&(1-k)^{-1}\Big[K^{SS^\dagger}|D_SW|^2+K^{XX^\dagger}|D_X W|^2 \nonumber \\
&&+(K^{SX^\dagger} (D_S W) (D_X W)^\dagger +{\rm h.c.})-3|W|^2\Big]
\eea
where $k\equiv \frac{1}{3}|S|^2+\frac{1}{3}|X|^2-\frac{1}{2}(\chi S^2+{\rm h.c.})$ and
\bea
D_S W&=& \lambda X S\bigg(1+\frac{|S|^2-3\chi S^2}{2(1-k_0-\frac{1}{3}|X|^2)}\bigg),\\
D_X W&=& \frac{1}{2}\lambda S^2 \bigg(1+\frac{|X|^2}{1-k_0-\frac{1}{3}|X|^2}\bigg).
\eea
By using the K\"ahler metric with $\gamma=0$ given in eqs.~(\ref{kahlerm}) and (\ref{inversekahlerm}),
the Jordan-frame scalar potential becomes
\bea
V_J&=&\frac{1-k_0}{1-k_0+\frac{1}{3}|S^\dagger-3\chi S|^2}\,|D_S W|^2+\frac{1-k+\frac{1}{3}|S^\dagger-3\chi S|^2}{1-k_0+\frac{1}{3}|S^\dagger-3\chi S|^2}\,|D_X W|^2 \nonumber \\
&&-\frac{1}{1-k_0+\frac{1}{3}|S^\dagger-3\chi S|^2}\cdot \frac{1}{3}X^\dagger (S-3\chi^\dagger S^\dagger)(D_S W)(D_X W)^\dagger+{\rm h.c.} \nonumber \\
&&-3(1-k)^{-1}|W|^2
\eea
with $k_0\equiv \frac{1}{3}|S|^2-\frac{1}{2}(\chi S^2+{\rm h.c.})$.
For $|X|\ll 1$ and $\chi|S|^2\gg 1$, we obtain the Jordan-frame scalar potential as
\be
V_J\simeq \frac{1}{4}|\lambda|^2 |S|^4 -\frac{|\lambda|^2}{6\chi}|X|^2 (S^2 +S^{\dagger 2})+{\cal O}\Big(\frac{|\lambda|^2}{\chi^2}|X|^4\Big).
\ee
Thus, the Jordan-frame scalar potential is unstable in the direction of the real part of the $S$ singlet.
Consequently, from eq.~(\ref{jspot}), with $S\equiv |S|e^{i\theta}$, the Einstein-frame scalar potential is
\bea
V_E&=&(1-k)^{-2} V_J \nonumber \\
&\simeq & \frac{|\lambda|^2M^4_P}{4\chi^2\cos^2(2\theta)}\bigg[1-\frac{2M^2_P}{\chi|S|^2\cos 2\theta} 
+\frac{2}{3}\Big(\frac{1}{\cos 2\theta}-2\cos 2\theta\Big)\frac{|X|^2}{\chi|S|^2}\bigg]. \label{espot}
\eea
Thus, we find that the F-term contribution of the $X$ singlet approaches a positive constant for $\chi |S|^2\gg 1$ while the other terms are suppressed so there appears a vacuum energy required for the inflation.
Here we have recovered the Planck scale, $M_P$.

Then, minimizing the potential for the angle at $\theta\simeq 0$ from the first term in Eq.~(\ref{espot}),
the scalar potential becomes
\be
V_E\simeq \frac{|\lambda|^2M^4_P}{4\chi^2}\bigg[1-\frac{2M^2_P}{\chi|S|^2}-\frac{2|X|^2}{3\chi|S|^2}\bigg].
\ee
Therefore, the slow-roll inflation along the $S$ singlet is possible for $\chi|S|^2\gg 1$ so the Hubble scale during the inflation is given by $H^2\simeq \frac{V_E}{3M^2_P}\simeq \frac{|\lambda|^2M^2_P}{12\chi^2}$.
However, even in the Einstein frame, we get a tachyonic effective mass of the $X$ singlet, ending up with the instability of this singlet direction.  

On the other hand, the kinetic terms for the singlets in the Einstein frame are given by
\be
{\cal L}_{\rm kin}\simeq -\frac{3M^2_P}{|S|^2}|\partial_\mu S|^2 -\frac{M^2_P}{\chi |S|^2}|\partial_\mu X|^2-\Big(\frac{M^2_P XS}{\chi|S|^4}\partial_\mu S \partial^\mu X^\dagger +{\rm h.c.}\Big). \label{einskin}
\ee
Then, at the minimum with $\theta\simeq 0$, in terms of the canonical inflaton field, we obtain the Lagrangian density as
\bea
{\cal L}_{\varphi,X}&\simeq& -\frac{1}{2}(\partial_\mu\varphi)^2-e^{-2\varphi/(\sqrt{6}M_P)}|\partial_\mu X|^2
-\frac{1}{\sqrt{6}M_P}e^{-2\varphi/(\sqrt{6}M_P)}\partial_\mu\varphi \partial^\mu |X|^2 \nonumber \\
&&-\frac{|\lambda|^2M^4_P}{4\chi^2}\bigg[1-2 e^{-2\varphi/(\sqrt{6}M_P)}-\frac{2}{3}e^{-2\varphi/(\sqrt{6}M_P)}\frac{|X|^2}{M^2_P}\bigg]
\eea
with $\varphi\equiv\frac{\sqrt{6}}{2}M_P\ln(\chi|S|^2/M^2_P)$.
Consequently, we find that in the canonical field basis, 
the $X$ singlet has a tachyonic effective mass of order the Hubble scale\footnote{For the approximate dS background with a slow-rolling $\varphi$, and $|X|\ll 1$, the equation of motion for $X$ is $\ddot{X}+3H{\dot X}\simeq -m^2_X X$. So, for $m^2_X\simeq-2H^2$, the small perturbation of the $X$ singlet would grow exponentially as $X\propto {\rm exp}((\sqrt{17}-3)Ht/2)$ during inflation. Since the time scale of exponential growth is $t_X\simeq \frac{1.78}{H}$ so it would make the slow-roll inflation with 60 efoldings impossible. I would like to thank M. Giovannini for pointing this out.}
\be
m^2_X\simeq K^{X X^\dagger} V_{E,X X^\dagger} \simeq -\frac{|\lambda|^2M^2_P}{6\chi^2}\simeq -2H^2.
\ee 
Therefore, the $X$ singlet would roll down fast into a minimum of the full scalar potential, dominating the scalar field
dynamics and spoiling the slow-roll inflation along the $S$ singlet.  
This problem seems generic for the canonical scalar kinetic terms in the Jordan frame supergravity\footnote{One can compare this case to the generalized chaotic inflation for the minimal K\"ahler potential and the superpotential $W=XS^n$  with $n$ being a natural number in Refs.~\cite{chaoticsugra,westp} where the accompanying singlet has a vanishing mass at the origin.}.

\section{A solution to the tachyonic mass problem}

In this section, we propose a simple solution to the tachyonic mass problem encountered in the inflation model of Jordan frame supergravity.
For this, we introduce a higher order correction with the $X$ singlet to the frame function as follows,
\be
\Delta \Omega = -\gamma (X^\dagger X)^2 \label{add}
\ee
where $\gamma=c\frac{M^2_P}{M^2}$ with $c$ being a dimensionless parameter and $M$ being the mass of heavy fields
that are integrated out. Here we expressed the coefficient $\gamma$ as being dimensionless in units of $M_P=1$.
In the presence of the higher order correction (\ref{add}), 
the Jordan-frame kinetic term for the $X$ singlet becomes non-canonical.
However, we keep the Jordan-frame kinetic term for the $S$ singlet to be canonical.
In this case, the additional non-canonical kinetic terms coming from $b_\mu$ in eq.~(\ref{auxvec}) still vanish for the frozen angular modes.

Generically, however,  both $(X^\dagger X)(S^\dagger S)$ and $(S^\dagger S)^2$ terms could be also generated after the heavy fields are integrated out. 
The $(X^\dagger X)(S^\dagger S)$ term corresponds to the effective wave function renormalization of the $X$ singlet during the inflation driven at a nonzero $|S|$. 
So, it does not affect much either the inflation dynamics or the $X$ singlet.
However, in order to maintain the behavior of the chaotic inflation with the non-minimal coupling at a large $|S|$, 
the $(S^\dagger S)^2$ term must be suppressed compared to the non-minimal coupling term. 
That is, for $\gamma_s|S|^4\ll \chi |S|^2$ with $\gamma_s$ being the coefficient of the $(S^\dagger S)^2$ term, together with the chaotic inflation condition, $\chi |S|^2\gg 1$, we need $\frac{1}{\chi}\ll |S|^2\ll \frac{\chi}{\gamma_s}$. Then, the resultant upper bound on the coupling is $\gamma_s\ll \chi^2$. For instance, integrating out heavy fields of mass $M$, 
we would generate $\gamma_s= c_s\frac{M^2_P}{M^2}$, becoming $\gamma_s\sim c_s\chi^2$ for the heavy field mass saturating the unitarity bound as will be discussed in next section. Then, we would need $c_s\ll 1$ for the slow-roll inflation. 
As shown in the appendix B, 
the smallness of $c_s$ is guaranteed when the tree-level coupling of heavy fields to the $S$ singlet is forbidden by a $Z_2$ discrete symmetry. 
Even higher order non-holomorphic interactions for the $S$ singlet generated by the heavy fields
would be suppressed by the same $Z_2$ symmetry.
On the other hand, higher order corrections to the holomorphic part of the frame function will not be generated due to non-renormalization theorem \cite{higgsnmssm}.

Now we discuss the effect of the $(X^\dagger X)^2$ term on the tachyonic mass problem. 
Due to the correction term in the frame function, for $|X|\ll 1$ and $\chi|S|^2\gg 1$,
the Jordan-frame scalar potential is modified to
\bea
V_J&\simeq& \frac{1}{4}|\lambda|^2(1+4\gamma|X|^2)|S|^4-\frac{|\lambda|^2}{6\chi}|X|^2(S^2+S^{\dagger 2}) \nonumber \\
&&+\frac{\gamma |\lambda|^2}{\chi}|X|^4\bigg(\frac{4|S|^4}{S^2+S^{\dagger 2}}+\frac{1}{3}(S^2+S^{\dagger 2})\bigg)+{\cal O}\Big(\frac{|\lambda|^2}{\chi^2}|X|^4\Big).
\eea
Thus, the higher order correction in the frame function leads to a higher dimensional interaction term, $|X|^2|S|^4$, which gives rise to an additional effective mass for the $X$ singlet during inflation and overcomes the tachyonic instability.
Then, the Einstein-frame scalar potential is modified to
\bea
V_E\simeq\frac{|\lambda|^2M^4_P}{4\chi^2\cos^2(2\theta)}\bigg[1-\frac{2M^2_P}{\chi|S|^2\cos 2\theta} 
+4\gamma |X|^2+\frac{2}{3}\Big(\frac{1}{\cos 2\theta}-2\cos 2\theta\Big)\frac{|X|^2}{\chi|S|^2}\bigg].
\eea
Therefore, at the minimum with $\theta\simeq 0$, we get the resultant scalar potential in the Einstein frame,
\be
V_E\simeq \frac{|\lambda|^2M^4_P}{4\chi^2}\bigg[1-\frac{2M^2_P}{\chi |S|^2}+4\gamma \frac{|X|^2}{M^2_P}-\frac{2|X|^2}{3\chi|S|^2}\bigg].
\ee 
On the other hand, for $|X|\ll M_P$, the kinetic terms in the Einstein frame are the same as eq.~(\ref{einskin}).
Thus, with $\varphi=\frac{\sqrt{6}}{2}M_P\ln(\chi|S|^2/M^2_P)$, the Lagrangian density of the singlets is
\bea
{\cal L}_{\varphi,X}&\simeq& -\frac{1}{2}(\partial_\mu\varphi)^2-e^{-2\varphi/(\sqrt{6}M_P)}|\partial_\mu X|^2
-\frac{1}{\sqrt{6}M_P}e^{-2\varphi/(\sqrt{6}M_P)}\partial_\mu\varphi\partial^\mu |X|^2 \nonumber \\
&&-\frac{|\lambda|^2M^4_P}{4\chi^2}\bigg[1-2 e^{-2\varphi/\sqrt{6}}+\frac{2}{3}\Big(6\gamma -e^{-2\varphi/(\sqrt{6}M_P)}\Big)\frac{|X|^2}{M^2_P}\bigg]. \label{effaction}
\eea
So, the effective mass of the $X$ singlet becomes
\be
m^2_X\simeq K^{XX^\dagger}V_{E,XX^\dagger}\simeq \Big(12\gamma e^{2\varphi/(\sqrt{6}M_P)}-2\Big)H^2.
\ee
Consequently, we find that for $6\gamma e^{2\varphi/(\sqrt{6}M_P)}> 1$, 
the $X$ singlet can have a positive squared mass of order the Hubble scale during inflation. 
Then, the $X$ singlet can be stabilized at the origin and the slow-roll inflation is driven by the $S$ singlet.

Using $e^{-2\varphi/(\sqrt{6}M_P)}\sim 0.02$ for the correct spectral index as will be discussed in next section,
the tachyon-free condition becomes $\gamma> 0.003$. 
As shown in the appendix B, when heavy fields are coupled to the $X$ singlet but not to the $S$ singlet, they can be integrated out, generating the one-loop correction to the frame function. The resulting one-loop effective frame function has no higher order correction for the $S$ singlet but it contains the leading higher order term for the $X$ singlet as $\Delta\Omega=-\gamma(X^\dagger X)^2$ where $\gamma=c\frac{M^2_P}{M^2}$ with
$c=\frac{|\kappa|^4}{192\pi^2}$. Here $\kappa$ is a dimensionless coupling of the $X$ singlet to the heavy field. 
In this case, we need $c=\frac{|\kappa|^4}{192\pi^2}> 0.003\frac{M^2}{M^2_P}$ for the stable $X$ singlet during inflation. 
As will be discussed in next section, imposing the unitarity bound $M\sim M_P/\chi$,
the singlet coupling is constrained to be $|\kappa|>\frac{0.97}{\sqrt{\chi}}$.

\section{Observational constraints versus unitarity bound}

For the inflation model discussed in the previous section, we consider the observational consequences
and the unitarity bound on new physics scale.

First, from the Einstein-frame potential (\ref{effaction}) at $X=0$, the slow-roll parameters are determined as follows, 
\bea
\epsilon&\simeq& \frac{1}{2}\bigg(\frac{\frac{\partial V_E}{\partial \varphi}}{V_E}\bigg)^2\simeq \frac{4}{3} e^{-4\varphi_i/\sqrt{6}}, \\
\eta&\simeq& \frac{\frac{\partial^2 V_E}{\partial\varphi^2}}{V_E}\simeq -\frac{4}{3} e^{-2\varphi_i/\sqrt{6}}
\eea
where $\varphi_i\gg \sqrt{6}$ (or $|S_i|\gg \frac{1}{\sqrt{\chi}}$) is the scalar vev during inflation.
From the number of efoldings, $N\simeq 60$, we get the spectral index and the tensor to scalar ratio,
\be
n_s\simeq 0.968, \quad r\simeq 3.0\times 10^{-3}.
\ee

On the other hand, the density perturbation at horizon exit is given by
\be
\Delta^2_{\cal R}=\frac{V_E}{24\pi^2 M^4_P\epsilon}\simeq \frac{|\lambda|^2 N^2}{72\pi^2\chi^2}.
\ee
Thus, from the COBE normalization, $\delta_H=\frac{2}{5}\Delta_{\cal R}=(1.91\pm 0.17)\cdot 10^{-5}$,
we get a constraint on the ratio between the dimensionless inflation parameters, 
\be
\frac{\chi}{|\lambda|}\simeq 5\times 10^4.\label{constraintonchi}
\ee

The non-minimal coupling to gravity induces a new effective interaction between the graviton and the scalar field.
Then, the power counting for the scattering amplitude for the scalar field involving the effective interaction gives rise
to the unitarity bound on the maximum energy scale. In particular, the Hubble scale, which is the inflation energy scale,
must be much smaller than the unitarity bound such that the semi-classical approximation for inflation is justified.
In our case, the non-minimal coupling, $F=\chi S^2$, gives rise to the effective interaction term in the Jordan frame,
\be
{\cal L}_{\rm eff}\simeq \Big(\frac{\chi}{M_P} S^2 + {\rm h.c.}\Big)\Box h^\mu_\mu \label{effint}
\ee
where $h^\mu_\mu$ is the trace part of the graviton.
Thus, from the power counting on scalar scattering \cite{unitarity,boundH,frameindep}, the upper bound allowed by unitarity on the new-physics scale is given by $\Lambda\simeq \frac{M_P}{\chi}$.
On the other hand, the Hubble scale during inflation is approximately given by $H\simeq \frac{|\lambda|M_P}{\chi}$.
For the semi-classical approximation for the inflation dynamics to be justified \cite{boundH}, we must have $H\ll \Lambda$, resulting in $|\lambda|\ll 1$. 
Suppose that $|\lambda|=0.01$. Then, from eq.~(\ref{constraintonchi}), we need to take the non-minimal coupling to be 
$\chi\simeq 5\times 10^2$. In this case, the quantum gravity scale becomes $\Lambda\simeq 0.01 M_P\sim 10^{16}$ GeV, being close to the GUT scale.

\section{Implications for the Higgs inflation in NMSSM}

The SM Higgs inflation with non-minimal coupling has been recently generalized to the supersymmetric case \cite{higgsnmssm} where there are two Higgs doublets required to be present for the anomaly cancellations.
In the MSSM, the Higgs quartic self-interaction comes from the gauge interactions, i.e. the D-term.
However, it turns out that the Higgs inflation does not work in the MSSM\footnote{The MSSM inflation with a flat direction such as $\phi=LLe$ or $udd$ can occur at the inflection point, which requires a fine-tuning between soft mass parameters for the flat direction \cite{mazumdar}.} because the slow-roll conditions are not satisfied along the $\tan\beta$ direction for a nonzero D-term \cite{higgsnmssm}. Therefore, we need an additional quartic self-interaction for the Higgs from the F-term.

The extension of the MSSM with a gauge singlet has been considered for solving the $\mu$ problem \cite{nmssm}.
In the NMSSM, the same term giving rise to the $\mu$ term in the superpotential leads to an additional quartic self-interaction for the Higgs in the scalar potential. 
The NMSSM extension of the Higgs inflation \cite{higgsnmssm} 
has been proposed with the following frame function and the superpotential,
\bea
\Omega&=&-3+H^\dagger_u H_u + H^\dagger_d H_d + X^\dagger X +\frac{3}{2}(\chi H_u H_d +{\rm h.c.}), \label{framenmssm} \\
W&=& \frac{1}{2}\lambda X H_u H_d +\frac{1}{3}\rho X^3
\eea
where $H_u,H_d$ are the Higgs doublets, $X$ is the SM singlet and $\chi,\lambda,\rho$ are dimensionless parameters.
In this case, the frame function or the K\"ahler potential in this model has a particular Higgs-dependent structure due to the non-minimal coupling so the $\mu$ term of order the SUSY breaking scale can actually be generated within supergravity \cite{gmmech} as will be discussed later in this section.
This model with $\beta=\frac{\pi}{4}$ where $\tan\beta\equiv \frac{\langle H_u\rangle}{\langle H_d\rangle}$, i.e. the D-flat direction, is similar to our model with two singlets in Jordan frame supergravity proposed in this paper.
That is, we can identify the inflaton $S$ in our toy model with the Higgs doublets satisfying the D-flat condition. 
In this NMSSM extension, it has been shown \cite{jsugra} that the inflationary trajectory with $X=0$ in this model has a tachyonic instability in the same way as analyzed in section 3. 

In order to solve the problem with a tachyonic mass of order the Hubble scale, we can add the same higher order term, $\Delta\Omega=-\gamma (X^\dagger X)^2$, for the $X$ singlet in the frame function (\ref{framenmssm}) as in Section 4.
In the NMSSM, however, there is an additional tachyonic mass for the $X$ singlet coming from the cubic term in the superpotential \cite{jsugra}: $\Delta m^2_X\simeq -\frac{|\lambda\rho|M^2_P}{\chi}\simeq -\frac{12\chi|\rho|}{|\lambda|}H^2$. 
Thus, for $\Delta m^2_X\gtrsim -H^2$, from $\frac{\chi}{|\lambda|}\simeq 5\times 10^4$, we need to choose a very small cubic coupling, $|\rho|\lesssim 10^{-5}$.
Therefore, if $\gamma\gtrsim 0.003$, the higher order term in the frame function gives rise to the positive squared mass for the $X$ singlet of order the Hubble scale so the singlet cubic coupling in the superpotential tends to be disallowed by the PQ symmetry. In this case, since the non-minimal coupling breaks the PQ symmetry explicitly, there is no problem with a dangerous PQ axion. If we choose a larger value of the higher order term, $\gamma\gg 0.003$, then it is possible to allow for the sizable singlet cubic coupling. 

Now we are in a position to address the question on the fate of the trajectory $\beta=\frac{\pi}{4}$ at the end of inflation.
During the inflation, it has been shown \cite{jsugra} that the field $\beta$ rapidly approaches $\frac{\pi}{4}$ and stay there 
for $\chi (g^2+g^{'2})\gg \lambda^2$. For a large non-minimal coupling $\chi$ and a small $\lambda$, this condition is always satisfied. However, at the end of inflation, the stability of the trajectory $\beta=\frac{\pi}{4}$ depends on
whether $g^2,g^{'2}>2\lambda^2$ or not \cite{jsugra}. If the tachyonic instability is present at the end of inflation, the tachyonic preheating would lead to large fluctuations of the field $\beta$ and spontaneous symmetry breaking.
If $g^2\simeq g^{'2}\sim \frac{1}{2}$ is given by the GUT-scale values at the end of inflation and
the unitarity bound is satisfied for $|\lambda|\ll 1$, we get $g^2,g^{'2}\gg 2\lambda^2$ so there is no tachyonic instability
of the $\beta$ field at the end of inflation.

In the NMSSM, $\lambda$ can be small for the Hubble scale ($H\simeq \frac{\lambda M_P}{\chi}$) during inflation to be much lower than the unitarity cutoff ($\Lambda\simeq \frac{M_P}{\chi}$), without having a phenomenologically unacceptable light Higgs mass unlike the SM Higgs inflation.
Therefore, it will be interesting to investigate the phenomenological consequences of the reliable Higgs inflation on
the parameter space in the NMSSM.
In order to compare to the low-energy data, we need to consider the running of the coupling constants.
But, here we make a qualitative discussion assuming that the running coupling constants are not so significantly different from the ones during inflation.

If the Higgs doublets are the inflaton, a large non-minimal coupling must be introduced in the frame function, generating the additional contribution to the $\mu$ term by supergravity effect.
That is, the effective $\mu$ term is given by the addition of the non-minimal coupling and the superpotential term as follows,
\be
\mu = \frac{3}{2}\chi m_{3/2} + \frac{1}{2}\lambda \langle X\rangle
\ee
where $m_{3/2}=|\langle e^{K/2}  W\rangle|$ is the gravitino mass.
In order for the $\mu$ term to be of the soft mass scale for electroweak symmetry breaking, we have to get $m_{3/2}\sim \frac{m_{\rm soft}}{\chi}$ and $\langle X\rangle\leq\frac{m_{\rm soft}}{\lambda}$.
Suppose that $\lambda\sim 0.01$ and $\chi\sim 10^2$, satisfying the constraints coming from the COBE normalization (\ref{constraintonchi}) and the unitarity bound on the Hubble scale discussed below Eq.~(\ref{effint}). Then, for $m_{\rm soft}\sim 1$ TeV, we would need the gravitino mass to be $m_{3/2}\sim 10$ GeV while the $X$ singlet VEV is to be $\langle X\rangle \leq 100$ TeV. 
In order to get such a small gravitino mass, gauge mediation must be dominant over gravity mediation.
In this case, when $R$-parity is conserved, the gravitino is LSP and can be either a non-thermal dark matter with neutralino NLSP \cite{feng} or a thermal dark matter for the reheating temperature $T_R\sim 10^8{\rm GeV}$ \cite{steffen}.

Here a comment on the Higgs physics is in order. 
Since the $\lambda$ coupling is so small, the tree-level contribution to the Higgs mass
coming from $\lambda$ is suppressed. Furthermore, the mixing between the singlet and the neutral component of the MSSM Higgs doublets is small so the lightest neutral Higgs in this NMSSM should be of the MSSM type.

\section{Conclusion}

We have reconsidered the inflationary model with a large non-minimal coupling in supergravity
and have shown that for the minimal Jordan frame function, the non-inflaton field gets a tachyonic mass of order the Hubble scale during inflation, developing the instability of the slow-roll inflation. We have shown that the tachyonic mass problem can be solved by introducing a higher order correction in the frame function.
The necessary correction can be obtained after the heavy fields coupled only to the non-inflaton field are integrated out.

This result sheds light on the Higgs inflation in the NMSSM as the same tachyonic mass problem of the singlet is solved by a similar higher order correction in the frame function.
Moreover, when the singlet coupling to the Higgs doublets in the NMSSM is made small, the Higgs inflation is a viable possibility within the semi-classical approximation in the effective theory even with a large non-minimal coupling.
Thus, combining both the observational constraints on the inflation with the unitarity bound on the new physics,
we found some interesting consequences on the NMSSM phenomenology.
First, the large non-minimal coupling generates the $\mu$ term which is much larger than gravitino mass.
Thus, gravitino becomes a dark matter candidate. In this case, one has to explain how the soft mass parameters of order the $\mu$ term can be much larger than gravitino mass for electroweak symmetry breaking.
When gauge mediation is dominant over gravity mediation, it is possible to have $m_{3/2}\ll m_{\rm soft}$.  
Second, due to a necessary small singlet coupling to the Higgs doublets, the NMSSM Higgs looks more like the MSSM Higgs.

In order to make sure of the naturalness of the Higgs inflation with a large non-minimal coupling in supergravity, one should also take into account the loop corrections of the inflaton potential due to the spontaneous SUSY breaking during inflation.
We leave this important question in a future work.

\section*{Acknowledgments}
The author thanks Jim Cline for his interest and encouragement on the work
and Jose Espinosa, Massimo Giovannini and Graham Ross for comments and discussions.

\def\theequation{A.\arabic{equation}}
\setcounter{equation}{0}
\vskip0.8cm
\noindent
{\Large \bf Appendix A: The K\"ahler metric}
\vskip0.4cm
\noindent

The K\"ahler metric $K_{i{\bar j}}=\partial_i\partial_{\bar j}K$ ($\phi^i=S,X$) for $K=-3\ln(1-k(S,X,S^\dagger,X^\dagger))$ 
is given by
\be
K_{i{\bar j}}=\frac{3}{(1-k)^2}\left(
\begin{array}{ll}
(1-k)k_{S{\bar S}}+|k_S|^2  & (1-k)k_{S{\bar X}}+k_S k_{\bar X} \\
(1-k)k_{{\bar S}X}+k_{\bar S}k_X  & (1-k)k_{X{\bar X}}+|k_{X}|^2 
\end{array}\right).
\ee
On the other hand, the first derivative of the K\"ahler metric is $K_i=\frac{3k_i}{1-k}$.

The inverse K\"ahler metric is given by
\be
K^{i{\bar j}}=\frac{(1-k)^2}{3D}\left(
\begin{array}{ll}
(1-k)k_{X{\bar X}}+|k_X|^2  & -(1-k)k_{{\bar S}X}-k_{\bar S} k_{X} \\
-(1-k)k_{S{\bar X}}+k_{S}k_{\bar X}  & (1-k)k_{S{\bar S}}+|k_{S}|^2 
\end{array}\right)
\ee
with 
\be
D\equiv [(1-k)k_{S{\bar S}}+|k_{S}|^2][(1-k)k_{X{\bar X}}+|k_X|^2]
-|(1-k)k_{S{\bar X}}+k_S k_{\bar X}|^2.
\ee

For $k=\frac{1}{3}|S|^2+\frac{1}{3}(1-\gamma |X|^2)|X|^2-\frac{1}{2}(\chi S^2+{\rm h.c.})$ used in the text,
we get $k_{S{\bar X}}=k_{{\bar S}X}=0$. 
In this case, the K\"ahler metric and the inverse K\"ahler metric are given as follows,
\be
K_{i{\bar j}}=\frac{1}{(1-k)^2}\left(
\begin{array}{ll}
1-k+\frac{1}{3}|S^\dagger-3\chi S|^2  & \,\,\,\,\frac{1}{3}X(1-2\gamma|X|^2)(S^\dagger-3\chi S) \\
\frac{1}{3}X^\dagger (1-2\gamma |X|^2)(S-3\chi^\dagger S^\dagger)  & \,\,(1-k)(1-4\gamma|X|^2)+\frac{1}{3}|X|^2(1-2\gamma|X|^2)^2 
\end{array}\right),\label{kahlerm}
\ee
\be
K^{i{\bar j}}=\frac{(1-k)^2}{9D}\left(
\begin{array}{ll}
 (1-k)(1-4\gamma|X|^2)+\frac{1}{3}|X|^2(1-2\gamma|X|^2)^2 & -\frac{1}{3}X^\dagger(1-2\gamma|X|^2)(S-3\chi^\dagger S^\dagger) \\
-\frac{1}{3}X (1-2\gamma |X|^2)(S^\dagger-3\chi S)  &  1-k+\frac{1}{3}|S^\dagger-3\chi S|^2 
\end{array}\right) \label{inversekahlerm}
\ee
with
\bea
\frac{(1-k)^2}{9D}=(1-k)\bigg[(1-4\gamma |X|^2)\Big(1-k+\frac{1}{3}|S^\dagger -3\chi S|^2\Big) 
+\frac{1}{3}|X|^2(1-2\gamma|X|^2)^2\bigg]^{-1}.
\eea
The first derivatives of the K\"ahler metric are given by 
\be
K_S=\frac{S^\dagger-3\chi S}{1-k}, 
\quad K_X=\frac{X^\dagger(1-2\gamma|X|^2)}{1-k}.
\ee

\def\theequation{B.\arabic{equation}}
\setcounter{equation}{0}
\vskip0.8cm
\noindent
{\Large \bf Appendix B: One-loop frame function due to massive fields}
\vskip0.4cm
\noindent

We consider the one-loop K\"ahler correction coming from heavy fields coupled to the $X$ singlet.
We take a toy model providing the necessary higher order corrections to solve the tachyonic
mass problem in Jordan frame supergravity. 

We introduce two heavy chiral superfields, $\Phi_1$ and $\Phi_2$, which have $U(1)_G$ charges, $G[\Phi_1]=+1$ and $G[\Phi_2]=-1$. We impose a $Z_2$ symmetry to forbid the unwanted coupling to the $S$ singlet as follows,
\be
Z_2:\,\, \Phi_1\rightarrow -\Phi_1, \quad \Phi_2\rightarrow -\Phi_2, \quad S\rightarrow S, \quad X\rightarrow X.
\ee
Then, the additional superpotential for the heavy fields is given by
\be
W'=\frac{1}{2}\kappa X\Phi^2_1+M\Phi_1\Phi_2. \label{superpadd}
\ee
We assume that the additional heavy fields have canonical kinetic terms in the Jordan frame, i.e.
\be
\Omega_{\rm tree}=\Omega_0+\Phi^\dagger_1\Phi_1+\Phi^\dagger_2\Phi_2 \label{fullframe}
\ee
where $\Omega_0$ is the frame function given in Eq.~(\ref{framefunc}).

The general formula for the one-loop correction to the frame function in dimensional regularization(DR) \cite{1loopkahler} is 
\be
\Delta\Omega = -\frac{\Gamma(1-\frac{d}{2})}{2(4\pi)^{d/2}\mu^{d-4}}\sum_i\bigg[(1-6\xi_i)m^{d-2}_{B,i}+m^{d-2}_{F,i}-4m^{d-2}_{V_i}\bigg]
\ee
where $d=4-\epsilon$, $\mu$ is the renormalization scale in DR, $m_{B,i},m_{F,i},m_{V,i}$ are masses of real scalars, Weyl fermions and gauge bosons in the Jordan frame. 
Here $\xi_i$ are the non-minimal couplings of real scalars from ${\cal L}=-\sqrt{-g}\frac{1}{2}\sum_i\xi_i\phi^2_i R$.
We note that if the tree-level frame function
does not contain a holomorphic non-minimal coupling, its expansion should give rise to $\Phi^\dagger_i\Phi_i$ as the only leading term in the K\"ahler potential so scalar fields have a conformal coupling in the Jordan frame \cite{1loopkahler}. 

In our case, since heavy scalar fields are conformally coupled to gravity with $\xi_1=\xi_2=\frac{1}{6}$, they do not contribute to the one-loop effective frame function. 
So, the one-loop frame function is given by the fermionic contribution only as follows,
\be
\Delta\Omega=-\frac{1}{32\pi^2}\sum_{i=1,2}\bigg(-\frac{2}{\epsilon}\,m^2_{F,i}+m^2_{F,i}\ln\Big(\frac{m^2_{F,i}}{\mu^2}\Big)\bigg). \label{1loopframe}
\ee
Since we are interested in the loop corrections to the $X$ singlet potential, we take $\langle \Phi_1\rangle=\langle \Phi_2\rangle=0$. So, from the superpotential (\ref{superpadd}), $\Phi_1,\Phi_2$ have no mixing with the $X$ singlet so they are decoupled. But, the eigenvalue masses of fermionic partners of $\Phi_1,\Phi_2$ depend on the VEV of the $X$ singlet and they are given by
\bea
M^2_{F,1,2}=M^2\Big(1+a\pm \frac{1}{2}\sqrt{2a+a^2}\Big), \quad a\equiv \frac{|\kappa X|^2}{2M^2}.\label{fmass}
\eea 
Therefore, after subtracting the divergences in DR, we obtain the renormalized one-loop frame function as
\bea
\Delta\Omega&=&-\frac{1}{32\pi^2}\bigg[2M^2\ln\Big(\frac{M^2}{\mu^2}\Big)+\Big\{\ln\Big(\frac{M^2}{\mu^2}\Big)+2\Big\}|\kappa X|^2 +\frac{|\kappa X|^4}{6M^2}\bigg].
\eea
The first term corresponds to the renormalization of the Newton constant and the second term is the wave function renormalization of the $X$ singlet while the last term is the higher order interaction term.

\end{document}